\documentclass[a4paper]{jpconf}
\usepackage{graphicx}
\usepackage{slashed}
\usepackage{amsmath}
\usepackage{amsfonts}
\usepackage{amssymb}
\usepackage{wrapfig}
\usepackage{hyperref}
\usepackage{color}
\usepackage{subfig}

\begin{document}


\title{The production of the heavy scalar $H$ in association with top and anti-top quarks}

\author{Skhathisomusa Mthembu$^{a,1}$, Mukesh Kumar$^{b, 2}$}
\address{$^{a}$ School of Physics, University of the Witwatersrand, Johannesburg, Wits 2050, South Africa.}  
\address{$^{b}$ National Institute for Theoretical Physics; School of Physics and Mandelstam Institute for 
Theoretical Physics, University of the Witwatersrand, Johannesburg, Wits 2050, South Africa.}
\ead{$^{1}$skhathisomusa.mthembu@cern.ch, $^{2}$mukesh.kumar@cern.ch}

\begin{abstract}
Following arXiv:1506.00612, an effective field theory approach has been introduced for two additional real
scalars, $H$ and $\chi$, to study the associated Higgs boson ($h$) production with top quarks at the Large
Hadron Collider with centre of mass energy $\sqrt s = 8$ TeV.
In all studies, one benchmark scenario is considered where the parameters of the model are constrained for
$m_H = 275$ GeV and $m_\chi = 60$ GeV.   
A comparative study has been performed for the production of the Higgs boson in association with a single top 
and top-pairs in the Standard Model with respect to the processes where $h$ is replaced with the heavy 
scalar $H$, which further decays via $H \to h \chi\chi$. 
The analysis has been performed at two stages where the top quarks are studied through a single-lepton 
channel firstly, and then in the second case with same-sign leptons, electrons or muons in di-lepton events.   
Different observables, like $b$ and $c$-tagged jet multiplicities, total missing energy, and
total scalar sum of transverse momenta of all jets and leptons, has been studied.   
\end{abstract}

\section{Introduction}
\label{intro}
In Ref.~\cite{vonBuddenbrock:2015ema}, detailed study have been presented to predict the transverse 
momentum ($p_T$) spectrum of the Higgs boson ($h$) following an effective field theory approach, where 
two additional real scalars $H$ and $\chi$ have been introduced to the Standard Model (SM) Lagrangian. 
The main process of study is $pp \to H \to h\chi\chi$, where $\chi$ is assumed to be a dark matter (DM) 
candidate - a source of missing energy ($E_T$).  The study follows from the results comprised of the 
measurements of the differential Higgs boson transverse momentum, double Higgs boson resonance 
searches, associated Higgs boson production with top quarks, and $VV$ resonances (where $V = W^\pm, Z$).
An efficient $\chi^2$ analysis has been performed in this study to constraint and fit the model parameters 
with respect to known experimental results. In conclusion the analysis fits the mass of the hypothetical real 
scalar boson, $H$, to be $m_H = 272^{+12}_{-9}$~GeV whereas the mass of $\chi$ is predicted around $60$~GeV.

In this study we take an opportunity to pick one of the benchmark scenarios from the above studies, where
$m_H = 275$ GeV and $m_\chi = 60$ GeV, to analyse the associated production of the Higgs boson with a 
single and a pair of top quarks. Section~\ref{hprod} deals briefly with the production modes of the Higgs boson 
at hadron colliders. The associated production of the Higgs with top quarks in the SM and beyond the SM, 
such as our additional scalars, are detailed in section~\ref{htop}, and in section~\ref{summ} we summarise our 
studies with concluding remarks.

\section{Higgs boson production at hadron colliders}
\label{hprod}
At hadron colliders, like the Large Hadron Collider (LHC), the Higgs boson production can be achieved via 
three main channels. Taking the descending order of magnitude of production cross sections, the modes are
(1) gluon-gluon fusion through top-quarks in one-loop diagrams, (2) vector-boson fusion through $W^\pm$
and Z-bosons and (3) in association with a pair of top-quarks and with a single top-quark. 
It is also known that the $h \to b\bar b$ final state is the dominant decay mode in the SM for $m_h = 125$~GeV, 
though has not yet been observed. The search for this decay mode is already precluded with overwhelming 
multijet backgrounds in the gluon fusion process (1), but the signal to background ratio for this decay mode can 
be improved via the modes (2) and (3)~\cite{Aad:2015gra}. 
Here we only focus on process (3) for our studies, which is advantageous for the direct measurement of 
top-Yukawa couplings, $y_t$. Since $y_t$ is close to unity (due to the large measured mass of the top quark),
it may also gain insight into the scale of new physics. 
This motivates our work on this production mode, not only in association with the $h$, but also with the 
comparatively heavy $H$ as discussed in the introduction.

\section{Associated Higgs production with top quarks}
\label{htop}
Before discussing our analyses for the top-quark associated $h$ and $H$ productions in detail, let us review 
the effective field theory - "a bottom-up" approach introduced in Ref.~\cite{vonBuddenbrock:2015ema} for the
scalars $H$ and $\chi$.   

For all numerical analyses the codes are written in {\texttt {Mathematica}}, {\texttt {C$^{++}$}}, {\texttt {Python}}, 
models are generated with {\texttt {FeynRules}} which are further used for event generation through 
{\texttt {MadGraph5}}. Events are then showered and hadronized using {\texttt {Pythia 8.2}}, plotting and analyses 
followed through the {\texttt {Rivet}} package.

\subsection{An effective Lagrangian approach}
\label{effth}
Considering two hypothetical real scalars, $H$ and $\chi$, an effective Lagrangian can be written with the 
following terms in addition to the SM Lagrangian:   
\begin{align}
\mathcal{L}_{H} &= -\frac{1}{4}~\beta_{g} \kappa_{_{hgg}}^{\text{SM}}~G_{\mu\nu}G^{\mu\nu}H
+\beta_{_V}\kappa_{_{hVV}}^{\text{SM}}~V_{\mu}V^{\mu}H,  \label{vh} \\
\mathcal{L}_{\text{Y}} &= -\frac{1}{\sqrt{2}}~\Big[y_{_{ttH}}\bar{t} t H + y_{_{bbH}} \bar{b} b H\Big],\\ 
\mathcal{L}_{\text{T}} &=-\frac{1}{2}~v\Big[\lambda_{_{Hhh}}Hhh + \lambda_{_{h\chi\chi}}h\chi\chi + \lambda_{_{H\chi\chi}}H\chi\chi\Big], \\
\mathcal{L}_{\text{Q}} &= -\frac{1}{2}\lambda_{_{Hh\chi\chi}}Hh\chi\chi - \frac{1}{4} \lambda_{_{HHhh}}HHhh 
-\frac{1}{4}\lambda_{_{hh\chi\chi}}hh \chi\chi -\frac{1}{4} \lambda_{_{HH\chi\chi}}HH\chi\chi, \label{vq}
\end{align}     
where $\beta_g = y_{ttH}/y_{tth}$ is the scale factor with respect to the SM Yukawa top coupling, $y_{tth}$, and 
used to tune the effective $ggH$ coupling (as $\beta_V$ is used for $VVH$ couplings). 
This formalism considers $H$ production through the effective $ggH$ vertex, which further decays into $h$ with $\chi$.
The symmetry of the above Lagrangian follows from $\chi$ being stable and treated as a DM candidate.
For constraints on couplings, masses etc., we followed Ref.~\cite{vonBuddenbrock:2015ema} and selected a 
particular benchmark scenario where $m_H = 275$ GeV and $m_\chi = 60$ GeV for all analyses. 
\begin{table}
\begin{center}
    \begin{tabular}{l l c}
    \hline\hline
    \texttt{P.\#} 	&Process                                           &   Cross-section [pb] \\ \hline\hline
    \texttt{P.1}	&$p p \to t \bar t h$                              &   1.11 $\times \,10^{-1}$   \\
    \texttt{P.2}	&$p p \to t \bar t H$, $H \to h \chi \chi$         &   1.48 $\times \,10^{-5}$   \\
    \texttt{P.3}	&$p p \to t h j + \bar t h j$                      &   2.55 $\times \,10^{-1}$   \\
    \texttt{P.4}	&$p p \to t H j + \bar t H j$, $H \to h \chi \chi$ &   3.90 $\times \,10^{-3}$   \\\hline\hline
    \end{tabular}
\end{center}
\caption{Cross sections in pb for the mentioned processes are calculated with $m_t = 172$ GeV, 
$m_h = 125$ GeV, $m_H = 275$ GeV and $m_\chi = 60$ GeV at $\sqrt{s} = 8$ TeV using \texttt{cteq6l1} PDF sets. 
The renormalisation and factorisation scales for $t\bar t h (H)$ is taken to be $\mu_R = \mu_F = m_h (m_H)$, 
while for $t h (H)j + \bar t h (H)j$, $\mu_R = \mu_F = (m_t + m_h (m_H))/4$.}
\label{cs}
\end{table}
\begin{figure*}[!ht]
  \centering
  \subfloat[]{\includegraphics[width=0.32\textwidth,height=0.3\textwidth]{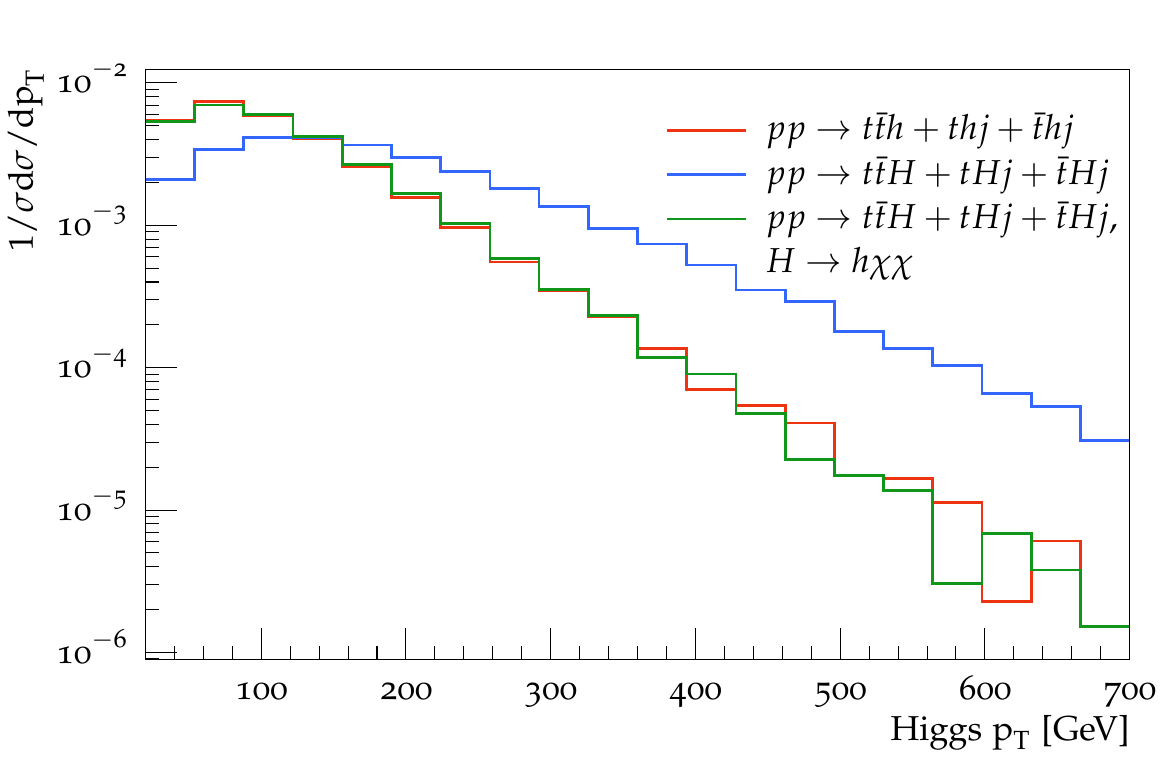}} 
  \subfloat[]{\includegraphics[width=0.32\textwidth,height=0.3\textwidth]{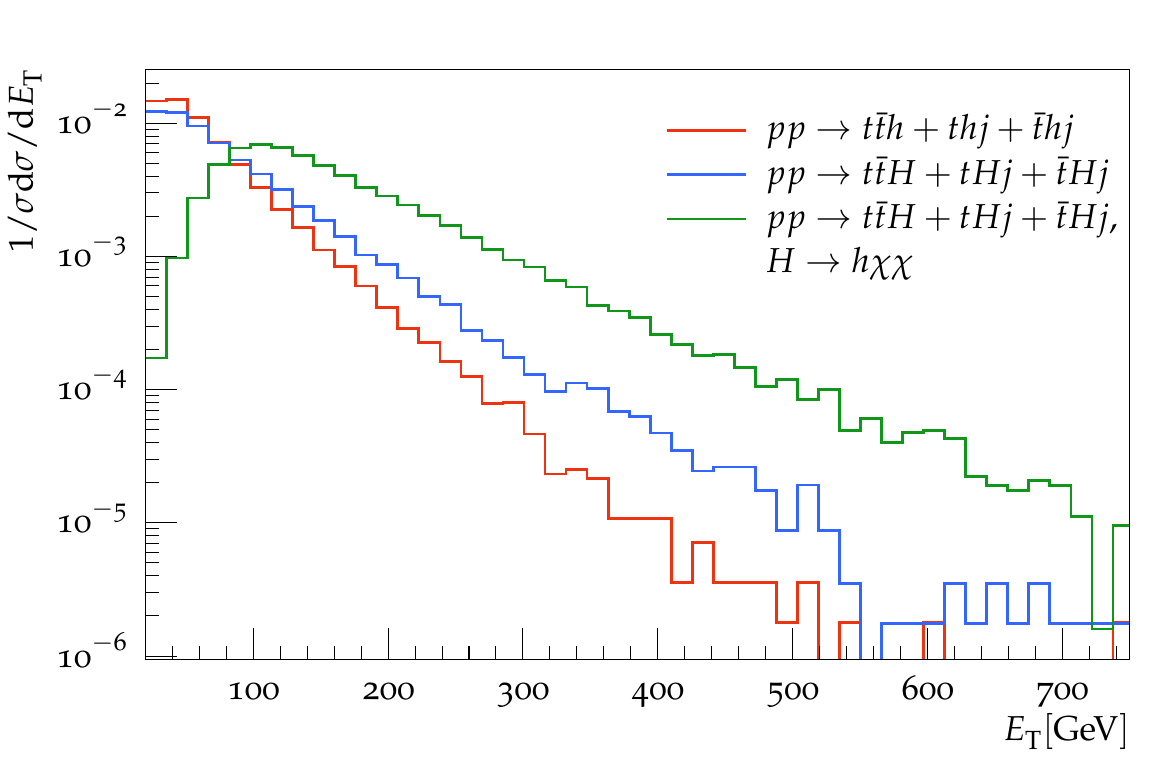}} 
  \subfloat[]{\includegraphics[width=0.32\textwidth,height=0.3\textwidth]{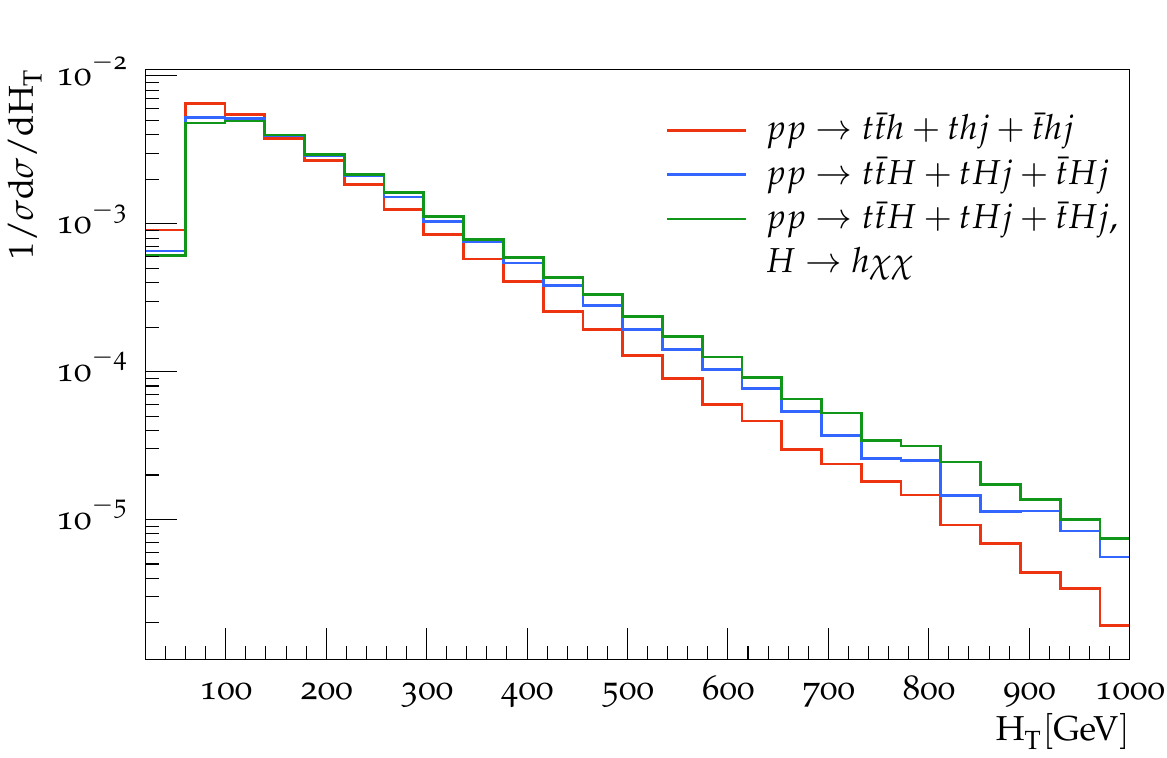}} 
  \caption{Differential distributions of (a) the Higgs $p_T$, (b) the total missing transverse momentum
  and (c) $H_T$, as discussed in the text for analyses \texttt{A.1}.}
 \label{fig:a}
\end{figure*}

\subsection{Higgs-boson production in association with a single and a pair of top quarks}
\label{pheno2hdm}
The processes for our studies are listed in Table~\ref{cs}, with the cross sections at leading order (LO) for 
corresponding benchmark choices, where events are generated using the \texttt{MadGraph5} package.
Here we shall notice that $h$ and $H$ production with a single top-quark is taken as only being in the
\texttt{5F}-scheme, i.e. considering only $2\to3$ processes with $b$-quark initiation and not in 
the \texttt{4F}-scheme\footnote{The \texttt{5F} scheme considers $b$-quark probability distribution 
functions (PDF) in the proton ($p$), whereas in the \texttt{4F} considers four flavours of quarks, along with $g$.}. 
The detailed studies for the schemes and production of the SM Higgs boson, in association with 
the single top quark at the LHC, at next-to-leading order (NLO) accuracy in QCD, can be found in 
Ref.~\cite{Demartin:2015uha}. 
A study on associated production of a top pair and a Higgs boson beyond NLO at the LHC can be found in
Ref.~\cite{Broggio:2015lya}. In Ref.~\cite{vonBuddenbrock:2015ema}, cross sections for these processes are
scaled-up with the factors given in Ref.~\cite{Heinemeyer:2013tqa}.

In Table~\ref{cs}, the SM production channels for a pair of top-quarks and a single top-quark with the Higgs
boson are represented as \texttt{P.1} and \texttt{P.3}, respectively. While the processes \texttt{P.2} and \texttt{P.4}
represents the associated $H$ production with a pair of top-quarks and a single top-quark, respectively, where
$H$ decays to $h\chi\chi$. The parameter choices for the processes \texttt{P.2} and \texttt{P.4} has been discussed
in the sub-section~\ref{effth}. Further we use the notation \texttt{P.13} and \texttt{P.24} as sum of the processes
\texttt{P.1}, \texttt{P.3} and \texttt{P.2}, \texttt{P.4}, respectively.\footnote{\texttt{P.13} and \texttt{P.24} refers 
inclusive productions of $h$ and $H$ in association with top-quarks, respectively.}  

It is known that in the SM a top quark decays almost exclusively to a $W^\pm$ boson and a $b$ quark. 
The $W^\pm$ boson further decays into a pair of leptons $\left( e\bar\nu_e, \mu\bar\nu_\mu, \tau\bar\nu_\tau \right)$,  
or into a pair of quark-jets. $\tau$ leptons produced by $W^\pm$ boson decays can also decay into leptons
$\left( e\bar\nu_e\nu_\tau, \mu\bar\nu_\mu\nu_\tau \right)$.
In our analyses, we followed the studies in two stages, by top-quark reconstructions
and event selections through:
\begin{itemize}
\item [\texttt{A.1}] A single lepton channel, and 
\item [\texttt{A.2}] Events with same-sign leptons, electrons ($e^\pm$) or muons ($\mu^\pm$) in di-lepton channels.
\end{itemize}
We follow some of the analyses for selection of events, cuts on $p_T$, (pseudo)rapidity ($\eta$),
etc. from Refs.~\cite{Aad:2015iha, Aad:2015gdg}.  

For analysis \texttt{A.1}, we select the events with jets and leptons $p_T > 25$ GeV within $|\eta| < 2.5$, and study various
kinematic distributions. Fig.~\ref{fig:a} shows the Higgs $p_T$, $E_T$ and $H_T$ distributions normalised with respect to
the total production cross sections of the corresponding processes. The two processes, i.e. the associated Higgs with a single 
top-quark and a pair top-quarks, are added together in all the plots. In this analysis, apart from the processes mentioned in
Table~\ref{cs}, we studied an additional process without decaying the heavy scalar $H$. The $p_T$ spectrum for $h$ 
in the processes \texttt{P.1-P.4}, and  the process where $H$ is not decaying, has different shapes in higher bins, as
shown in Fig.~\ref{fig:a}(a), since $m_H > 2 m_h$. However, the $E_T$ plot, Fig.~\ref{fig:a}(b), shows excesses of events
in higher bins for the processes where $H$ decays to $h\chi\chi$, since $\chi$ is also counted as a source of $E_T$. 
Fig.~\ref{fig:a}(c) shows the $H_T$ distribution, which are defined as scalar sums of all jets transverse momentum, $p_T^j$, 
defined as $H_T = \sum_{j} |\vec{p}^{\,j}_T|$, and small differences can be seen for the processes with respect to the SM.   

\begin{figure*}[!ht]
  \centering
  \subfloat[]{\includegraphics[width=0.4\textwidth,height=0.3\textwidth]{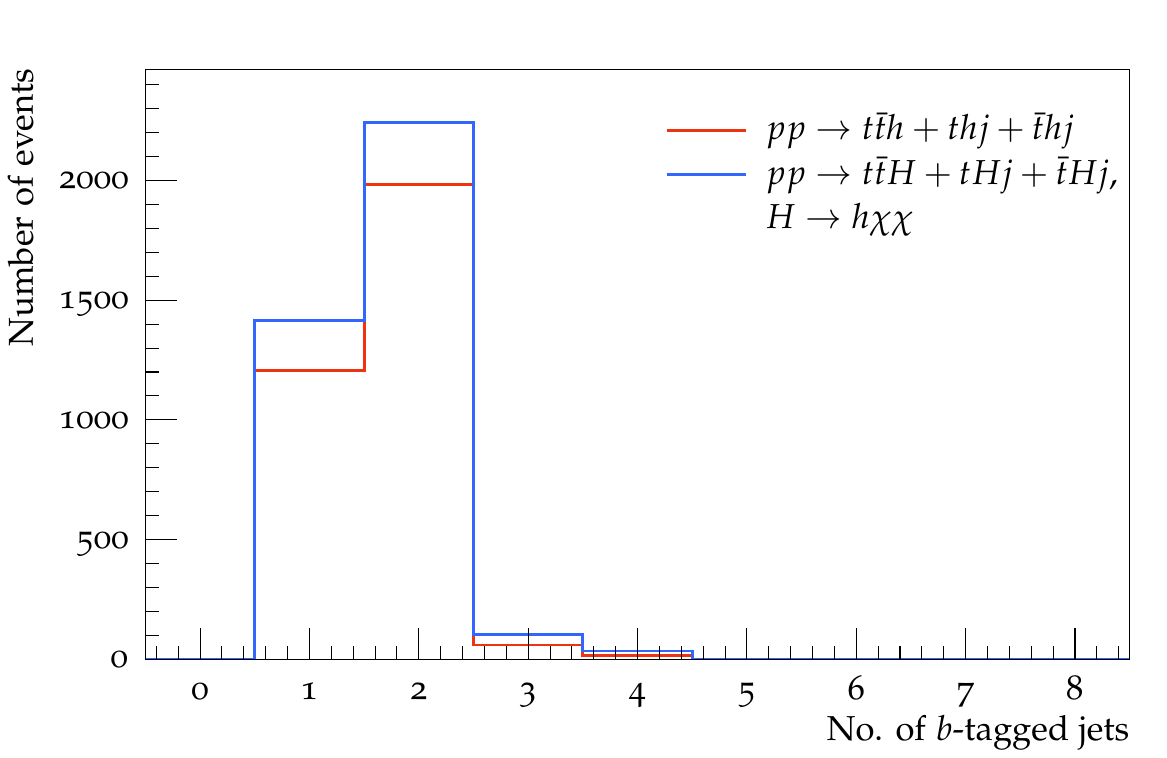}} \qquad
  \subfloat[]{\includegraphics[width=0.4\textwidth,height=0.3\textwidth]{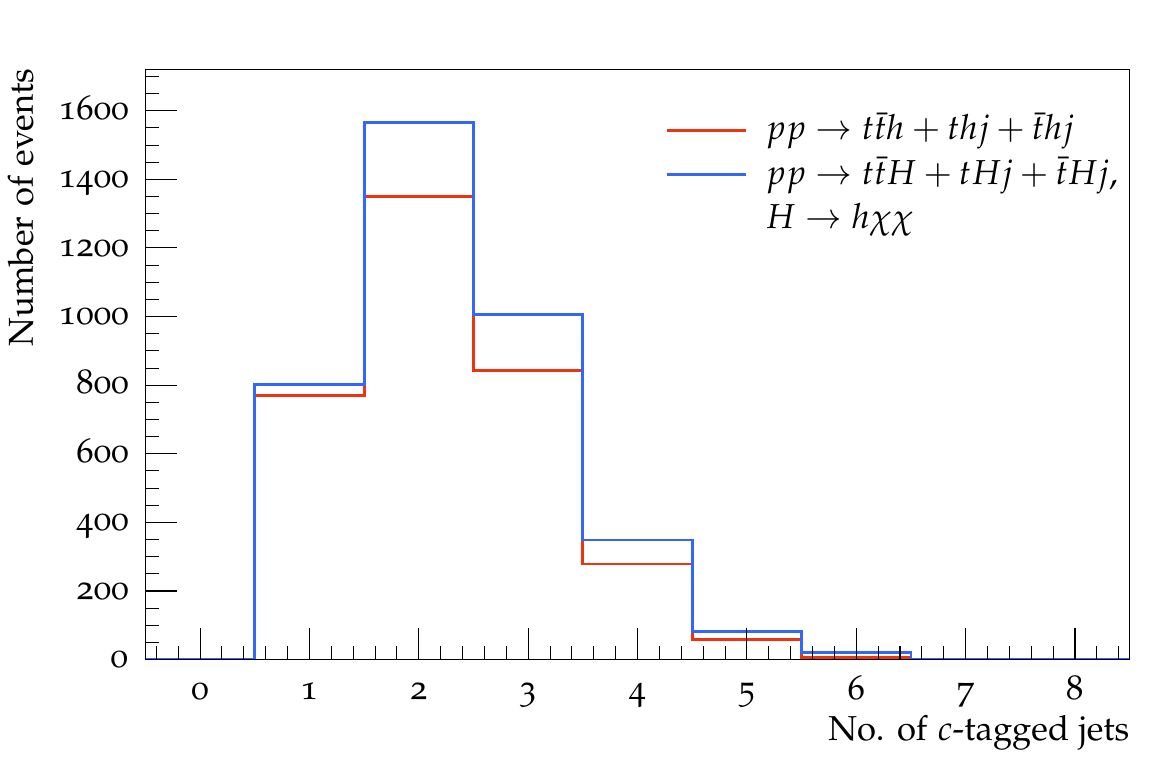}}
  \caption{Jet multiplicities of (a) $b$-tagged and (b) $c$-tagged jets, after preselections for the analyses \texttt{A.2}.}
 \label{fig:b}
\end{figure*}
\begin{figure*}[!ht]
  \centering
  \subfloat[]{\includegraphics[width=0.32\textwidth,height=0.3\textwidth]{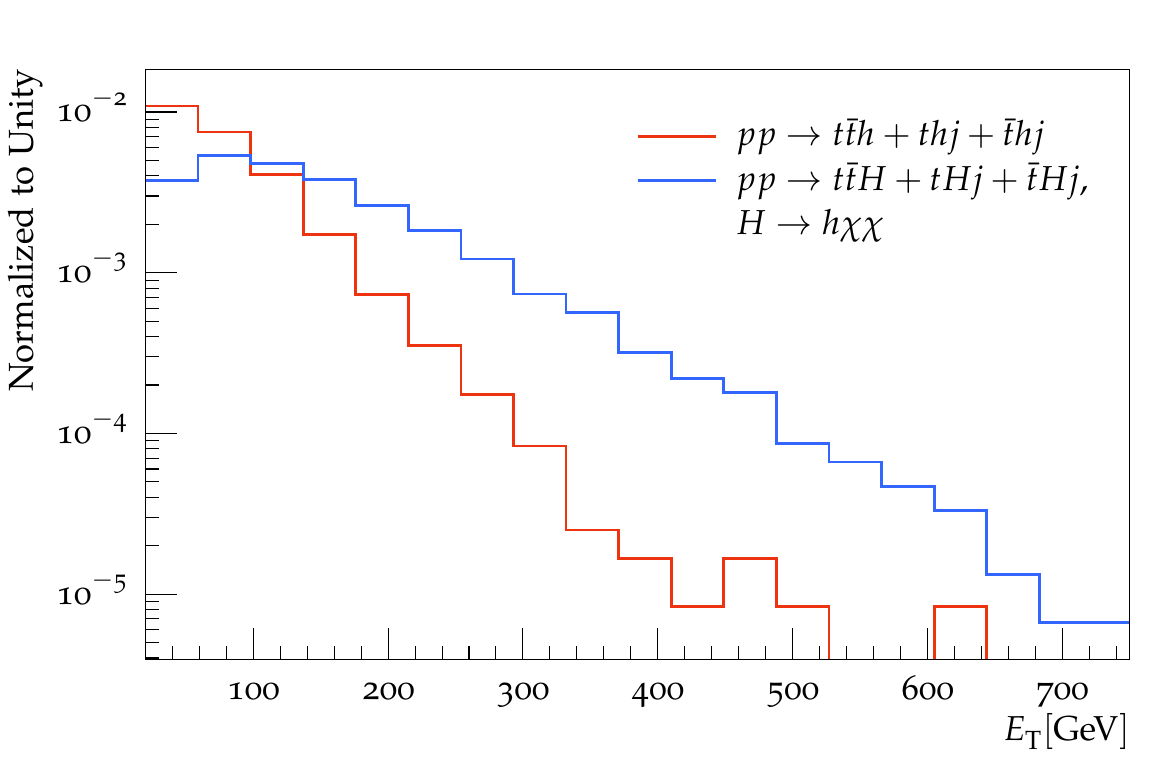}}
  \subfloat[]{\includegraphics[width=0.32\textwidth,height=0.3\textwidth]{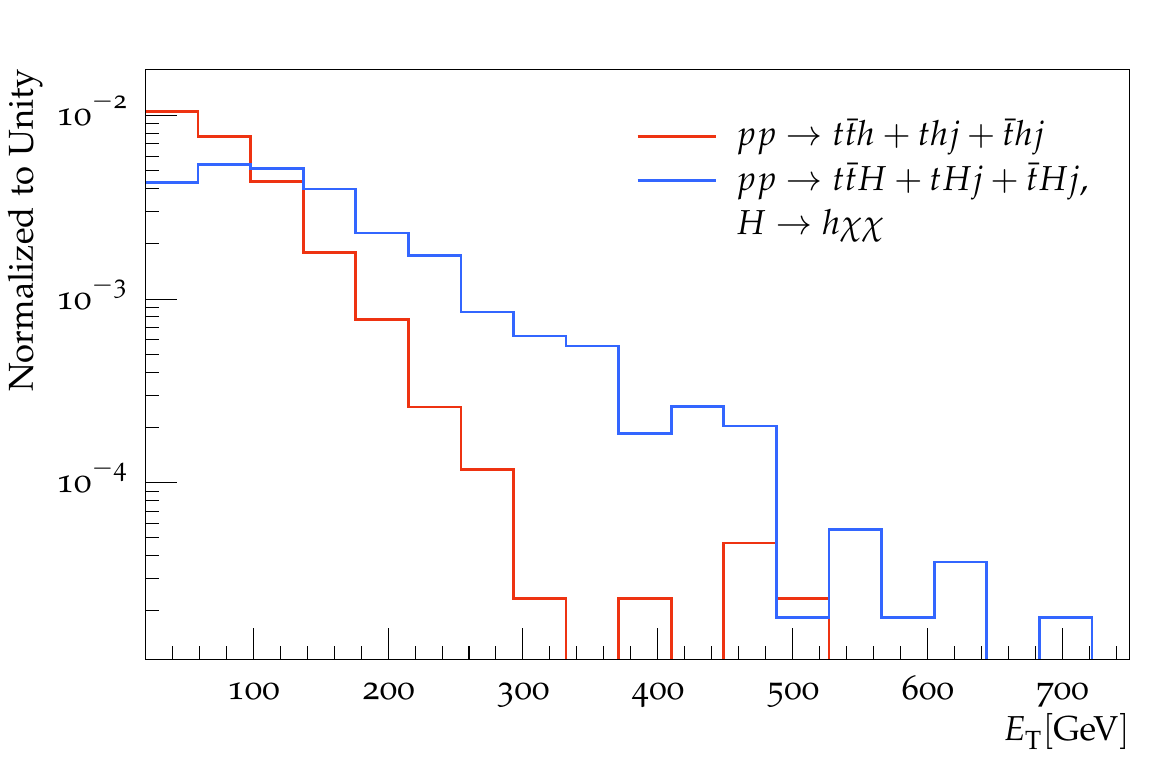}}
  \subfloat[]{\includegraphics[width=0.32\textwidth,height=0.3\textwidth]{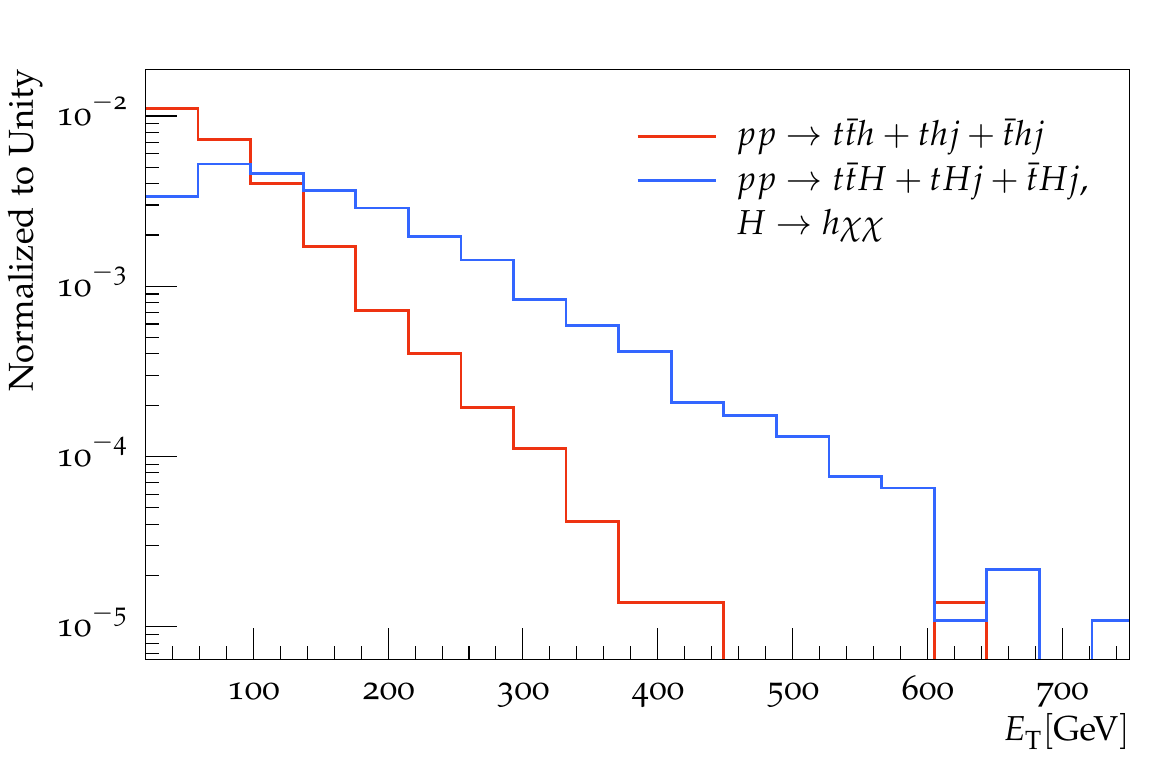}} \\
  \subfloat[]{\includegraphics[width=0.32\textwidth,height=0.3\textwidth]{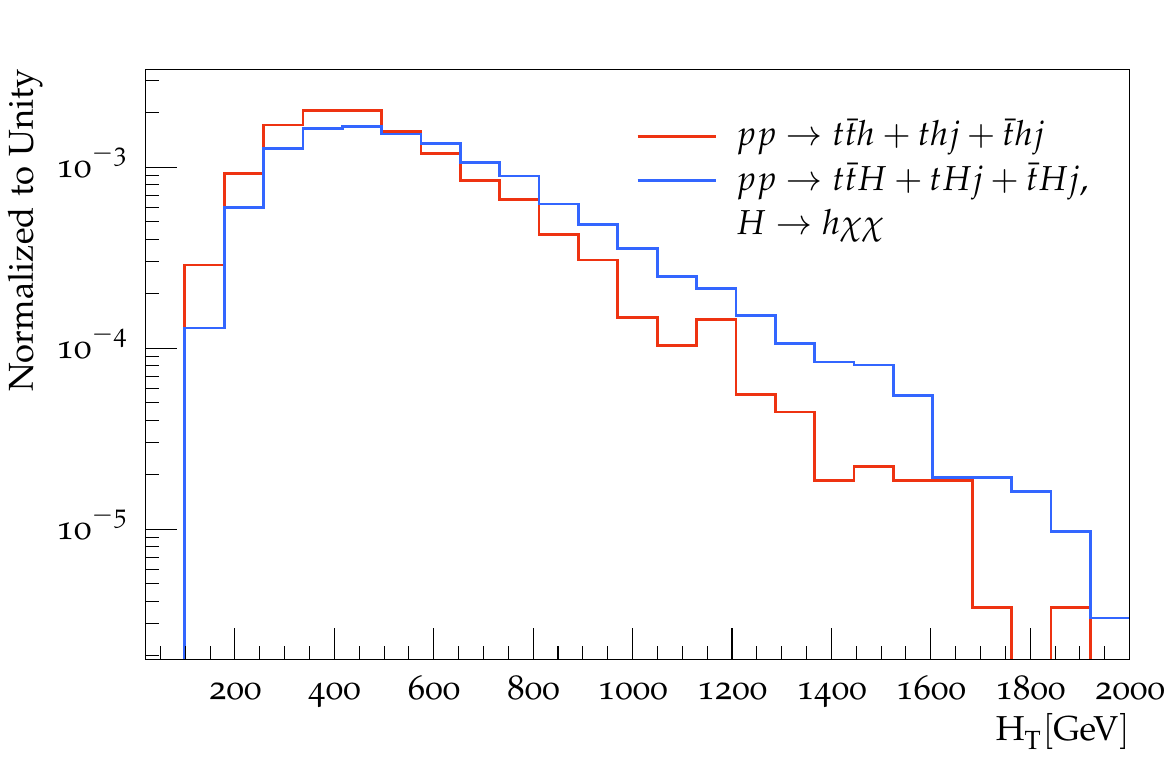}}
  \subfloat[]{\includegraphics[width=0.32\textwidth,height=0.3\textwidth]{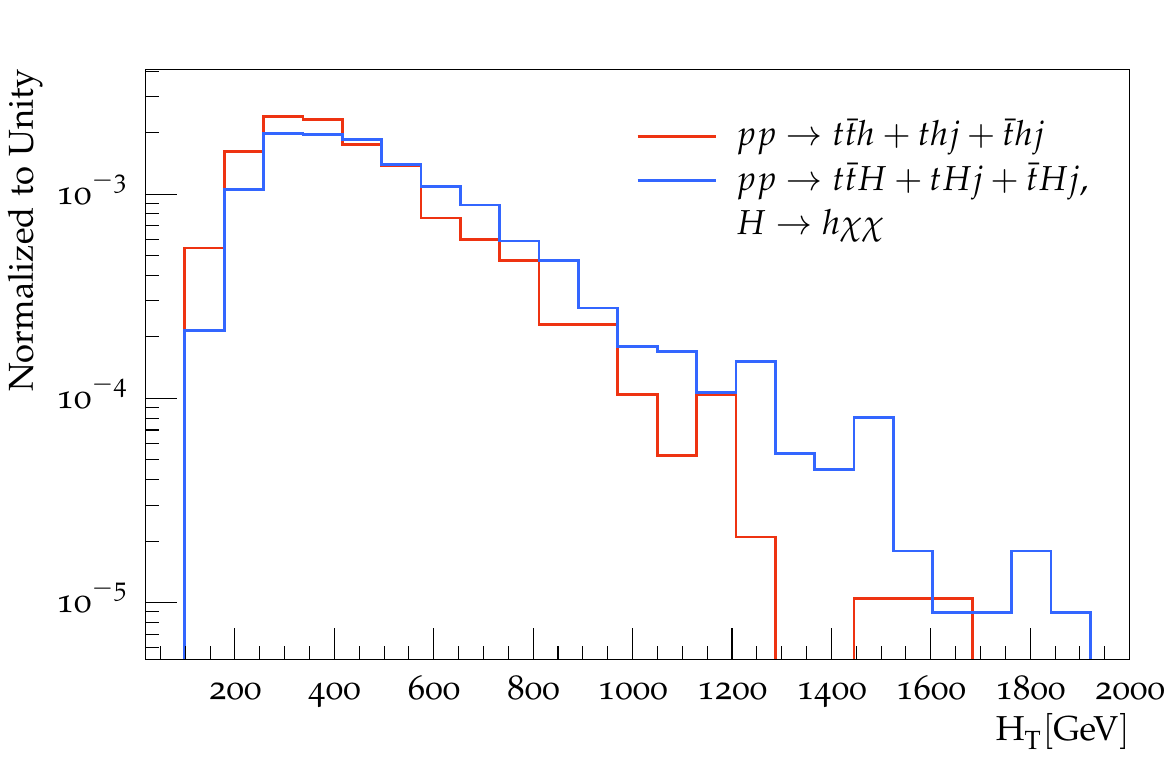}}
  \subfloat[]{\includegraphics[width=0.32\textwidth,height=0.3\textwidth]{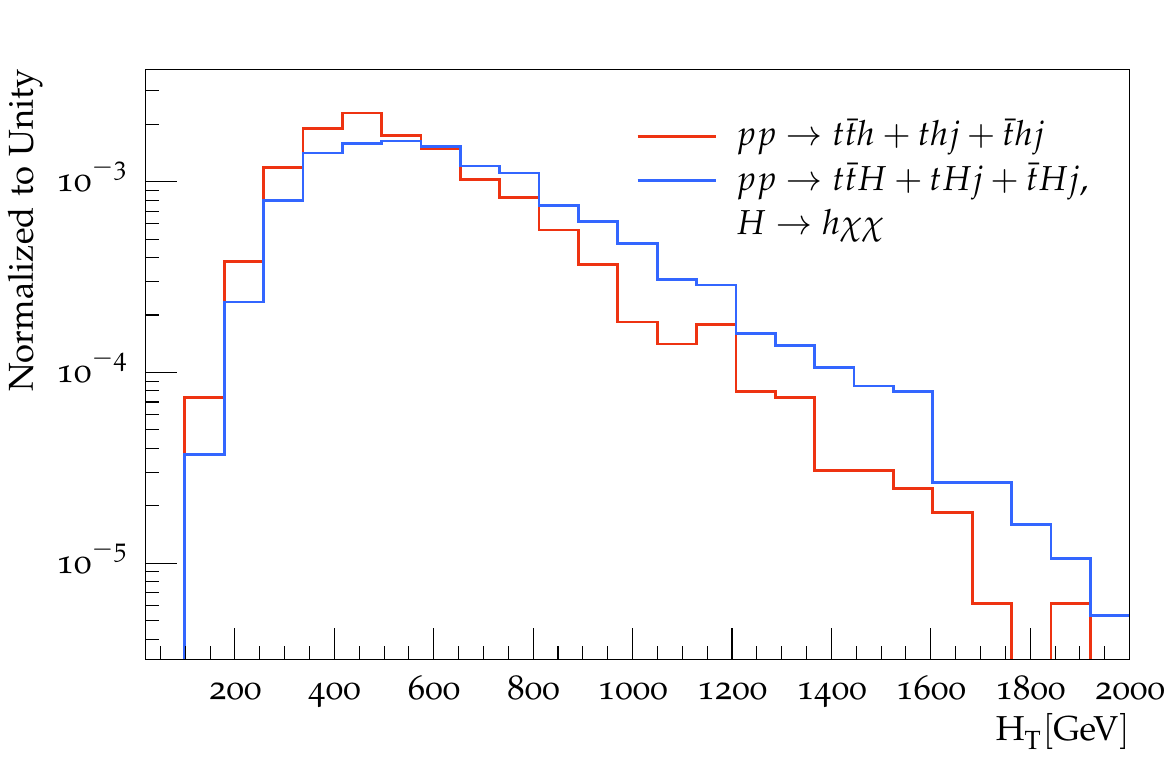}}   
  \caption{Distributions of the missing transverse momentum, $E_T$, and a scalar sum of jets
and lepton transverse momenta, $H_T$, after applying the preselection criteria for events respectively
with (a), (d) one $b$-tagged jets, (b), (e) two $b$-tagged jets and (c), (f) more than two $b$-tagged jets 
 (for analyses \texttt{A.2}).}
 \label{fig:c}
\end{figure*}

\begin{table}
\begin{center}
    \begin{tabular}{ c|p{5.8cm} |c c| c c| c c}    \hline\hline
    & Cut-flows  & \texttt{P.1}	& \texttt{P.2} & \texttt{P.3}	& \texttt{P.4} & \texttt{P.13}	& \texttt{P.24}   \\ \hline\hline
    (0) & Total events	& 30000	& 30000	& 30000	& 30000	& 60000	& 60000 \\
    (1) & 2 leptons with same sign	& 10933	& 11011	& 10023	& 10171	& 20956	& 21182 \\
    (2) & $|\eta| < 2.5$ for leptons	&  9614	&  9800	&  7360	&  7652	& 16974	& 17452 \\
    (3) & leading lepton
    $p_T > 25$ GeV, sub-leading lepton
    $p_T > 15$ GeV 	& 2305	& 2622	& 1156	& 1326	& 3461	& 3948 \\
    (4) & Number of jets $\geq 2$
    (at least 2 jets)	&  2301	&  2622	&  1115	&  1295	& 3461	& 3917 \\ \hline\hline
    (5) & exactly 1 $b$-tagged events	&   573	&   661	&   632	&   753	& 1205 	& 1414 \\ \hline\hline
    (6) & $\geq 2$ $b$-tagged events	&  1682	&  1921	&   377 	&   462 	& 2059  	& 2383 \\ \hline\hline
    & exactly 1 $b$-tagged events: &  &  &  &  &  &  \\
    (7) &    $400 < H_T < 700$, $E_T > 40$	& 189	& 291	& 116	& 265	& 305	& 556 \\
    (8) &    $H_T \geq 700$, $40 < E_T < 100$	&  26		&  22		&  16		&  11		&  42		&  33 \\
    (9) &   $700 \geq H_T$ , $E_T \geq 100$ 	&  36		& 184	&  21 	&  58 	&  57 	& 242 \\ \hline\hline
    & at least 2 $b$-tagged events: &  &  &  &  &  &  \\
    (10) &    $400 < H_T < 700$, $E_T > 40$	& 629 	& 829 	& 102 	& 198 	& 731	& 1027 \\
    (11) &    $HT \geq 700$, $40 < E_T < 100$	& 151 	& 134 	&  19 	&   8 		& 170 	& 142 \\
    (12) &    $700 \geq H_T$ , $E_T \geq 100$	& 143 	& 643 	&  16 	&  69 	& 159 	& 712 \\ \hline \hline
    (13) & $N_j\in$[2,4], $b$-tagged $\geq$ 1,
    $\Delta\phi_{ll} > 2.5$,
    $H_T > 450$, $E_T > 40$	& 33 		& 63 		& 32 		& 65 		& 65 		& 128 \\ \hline
    \hline
    \end{tabular}
\end{center}
\caption{Definitions of the different signal regions. $N_j$ is the number of jets that pass the
selection requirements, and $\Delta\phi_{ll}$ is the separation in $\phi$ between the leptons.
\texttt{P.13} and \texttt{P.24} are the sum of the processes
\texttt{P.1}, \texttt{P.3} and \texttt{P.2}, \texttt{P.4}, respectively.}
\label{all}
\end{table}

The analyses \texttt{A.2}, followed with preselection of events before any particular selections and are:
\begin{itemize}
\item [(1)] Selection of 2 leptons with same sign,
\item [(2)] $|\eta| < 2.5$ for leptons, 
\item [(3)] $p_T$ cut: leading lepton $p_T >$ 25~GeV, sub-leading lepton $p_T >$ 15~GeV, 
\item [(4)] Number of jets $\geq$ 2 (at least 2 jets) with $p_T > 25$~GeV within $|\eta| < 3$.
\end{itemize}
Thereafter special selection cuts are described in 
Table~\ref{all} with number of events for the processes shown in Table~\ref{cs}. 
Different differential distributions corresponding to these analyses are shown in Figs.~\ref{fig:b} and~\ref{fig:c}
after preselection criteria.
For these analyses we have defined $H_T$ as the scalar sum of jet and lepton transverse momenta: 
$H_T = \sum_{j,l} |\vec{p}^{\,j,l}_T|$. The $E_T$ distributions show an excess of events for $H \to h\chi\chi$ in 
comparison to the SM processes, since $\chi$ is the source of missing energy which was also discussed in the 
analyses \texttt{A.1}. Some excesses in the $H_T$ distributions, similar to $E_T$, is observed as shown in 
Figs.~\ref{fig:c}, where the distributions are obtained for a particular requirements on the number of $b$-tagged jets
in total events. 
The distributions in Figs.~\ref{fig:c} (a), (d) requires only one $b$-tagged jets,
while Figs.~\ref{fig:c} (b), (e) requires two $b$-tagged jets and Figs.~\ref{fig:c} (c), (f) requires more than two
 $b$-tagged jets among total events. 
  
In Table~\ref{all} different selection stages can be read after the horizontal line. The first block are all the preselection
criteria (from (1)-(4)), and then after each block (separated with a horizontal lines and numbered appropriately from (5)-(13)) 
follows a special selection in addition to the preselections of
events. Comparing \texttt{P.13} and \texttt{P.24}, which is basically the SM versus $H \to h\chi\chi$ process in last 
column of Table~\ref{all}, gives an idea of the number of events in both cases after each selection. In a few selection
choices the SM events are more than those beyond the SM, but in most of the cases beyond the SM events are higher than
the SM events. These analyses show how a heavy scalar $H$ production can be observed in experiments following the
decay $H \to h\chi\chi$ in top-quark associated Higgs boson productions at the LHC with $\sqrt s = 8$~TeV. 
        
\section{Summary and conclusions}
\label{summ}
Following Ref.~\cite{vonBuddenbrock:2015ema}, we discussed an effective field theory approach to add
two real scalars $H$ and $\chi$ in addition to the SM Lagrangian. Choosing one benchmark point
from this analysis, we studied the associated Higgs boson production with a single top-quark and a pair of
top-quarks in the SM, as well as beyond the SM, in particular $H \to h\chi\chi$ at the LHC with
$\sqrt s = 8$~TeV, $m_H = 275$~GeV and $m_\chi = 60$~GeV.

With two distinct stages of analysis, we studied these processes in detail with single lepton and
same-sign lepton, $e^\pm$ or $\mu^\pm$ di-lepton events. The study follows through analysing 
differential distributions of kinematics and then counting the number of events with special selection 
criteria on said events.
A comparative study for the SM and beyond the SM processes shows how the events for associated
Higgs production with top-quarks can be observed at the LHC with $\sqrt s = 8$~TeV. Any excesses
in the events at the LHC for the processes studied here, may be explained with the effective model
approach discussed with the introduction of $H$ and $\chi$, where in particular $H$ decays to $h\chi\chi$.
This study also leads to open a discussions of any possible new physics contributions for top-Yukawa 
couplings.     
  
\section*{Acknowledgements}
S.M. and M.K. would like to acknowledge Bruce Mellado, Alan S. Cornell, Deepak Kar and Stefan von 
Buddenbrock for fruitful discussions on the subject. S.M. is supported by funding from the National 
Research Foundation (NRF) of South Africa.

\end{document}